# MoHeat: A Modular Platform for High-Responsive Non-Contact Thermal Feedback Interactions

Jiayi Xu[1], Kazuma Nakamura[2], Yoshihiro Kuroda[3] and Masahiko Inami[1]

[1] Research Center for Advanced Science and Technology, The University of Tokyo, Tokyo, Japan

[2] Graduate School of Science and Technology, University of Tsukuba, Tsukuba, Japan

[3] Institute of Systems and Information Engineering, University of Tsukuba, Tsukuba, Japan

(Email: xu.jiayi@star.rcast.u-tokyo.ac.jp)

**Abstract ---** MoHeat is a modular hardware and software platform designed for rapid prototyping of highly responsive, non-contact thermal feedback interactions. In our previous work, we developed an intensity-adjustable, highly responsive, non-contact thermal feedback system by integrating the vortex effect and thermal radiation. In this study, we further enhanced the system by developing an authoring tool that allows users to freely adjust the intensity of thermal stimuli, the duration of stimuli, the delay time before stimuli, and the interval between alternating hot and cold stimuli. This modular approach enables countless combinations of non-contact thermal feedback experiences.

## 1 INTRODUCTION

Temperature plays a crucial role in the human sensory system, aiding not only in the recognition of material properties and the perception of environmental changes but also closely relating to emotions and pain [1]. Incorporating thermal feedback into interactive experiences is not merely a mimicry of reality; it holds the potential to regulate human emotions and create sensory experiences that resonate deeply with individuals.

While haptic modules have been extensively studied [2-6], research on thermal feedback is still in its early stages. Several studies have introduced thermal modules, such as TherModule [7] and PneuMod [8], which are compact and easily wearable on targeted skin areas. However, these devices are contact-based, relying on close proximity to deliver precise thermal feedback.

In our previous research, we developed an intensity-adjustable, high-responsive, non-contact thermal feedback system [9][10]. This system integrates cold airflow generated by the vortex effect and thermal radiation through visible light to provide both heating and cooling stimuli.

To further enhance and extend the capabilities of this system, this study introduces MoHeat, a modular platform. MoHeat offers a user-friendly authoring tool that enables interaction designers to customize and combine various intensities and durations of heating and cooling stimuli. This modular approach facilitates rapid iteration of thermal sensation prototypes, allowing for the creation of precise thermal feedback interactions.

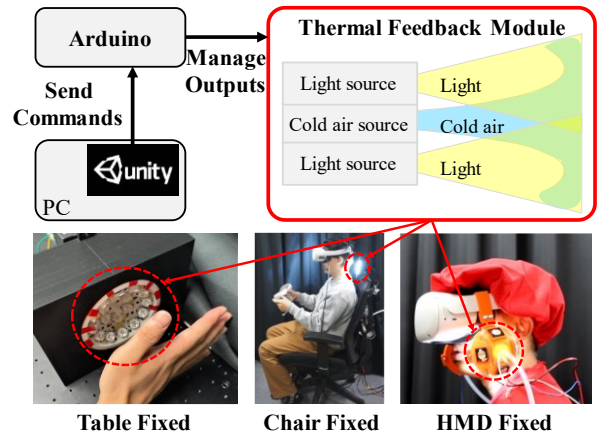

Fig. 1 Configuration of MoHeat

## 2 METHOD

MoHeat is capable of rapidly altering skin temperature, with a maximum heating rate of 0.6 °C/s and a minimum cooling rate of -0.3 °C/s, providing instantaneous thermal feedback [10]. As shown in Fig. 1, MoHeat utilizes a microcomputer (Arduino Uno) to manage the outputs of the thermal feedback system, with Unity sending commands to the microcomputer via serial communication. To streamline the development process and enhance usability, we have encapsulated the methods for both Arduino and Unity. Since the design of the module is compact, it can be customized for different use scenarios. Fig. 2 illustrates the authoring tool, which allows interaction designers to focus on creating interactive experiences without needing to delve into low-level coding details. The authoring tool primarily provides five types of stimuli patterns:

I. Cold: This pattern allows designers to customize the intensity, duration, and whether to introduce a delay before the cold stimuli are triggered.
II. Cold Level: This pattern offers five preset intensity levels for cold stimuli, with options to customize the duration and whether to introduce a delay before the stimuli are applied.
III. Hot: This pattern allows designers to customize the intensity, duration, and whether to introduce a delay before the hot stimuli are triggered.
IV. Hot Level: This pattern offers five preset intensity levels for hot stimuli, with options to customize the duration and whether to introduce a delay before the stimuli are applied.
V. Both Cold and Hot: This pattern enables designers to combine hot and cold stimuli to create unique thermal feedback experiences. Designers can independently set the intensity and duration of both stimuli, and as with other patterns, the overall duration and timing can be freely customized.

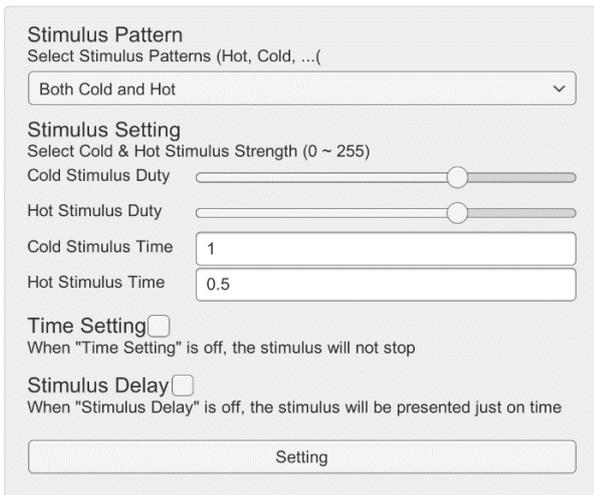

Fig. 2 Simple Configuration of Stimulus Patterns

## 3 DEMONSTRATION

Figures 3 through 5 show the examples of demonstrations developed using MoHeat. In these demonstrations, users can experience the thermal sensations of both global and local heat sources, such as the chill of moving through snowy mountains and the localized warmth of approaching flames. Users can also enjoy alternating cold and warm sensations synchronized with blue and red lighting.

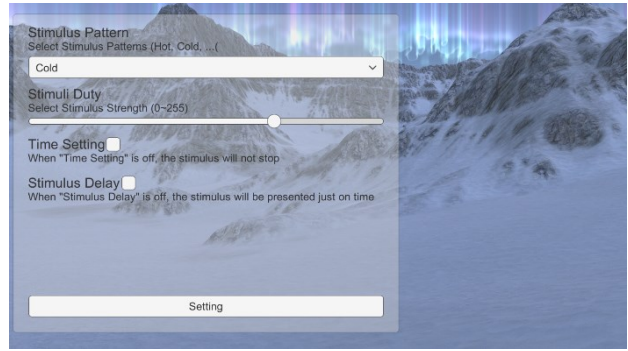

Fig. 3 Thermal Sensation of Global Heat Sources

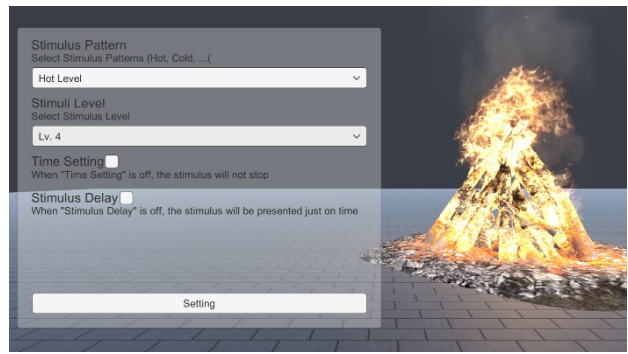

Fig. 4 Thermal Sensation of Local Heat Sources

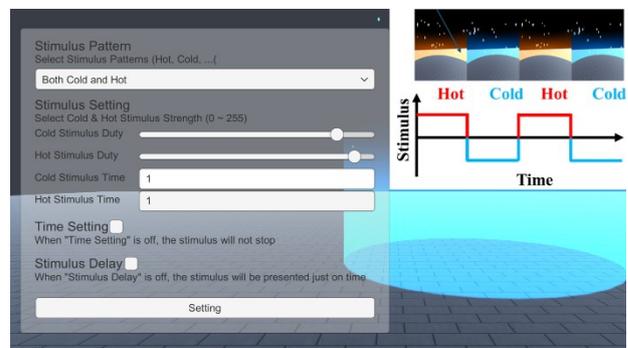

Fig. 5 Alternating Cold and Hot Sensations Over Time

## 4 CONCLUSION

In this study, we proposed MoHeat, a modular platform for high-responsive, non-contact thermal feedback interactions. This modular approach allows designers to customize and combine various intensities and durations of heating and cooling stimuli, enabling rapid iteration of thermal sensation prototypes until the desired feedback interaction is achieved. By providing a flexible and user-friendly platform, MoHeat opens new possibilities for creating immersive and emotionally engaging thermal feedback experiences. Our future research directions include reproducing real-world temperatures, identifying key components of thermal experiences, and automatically adjusting feedback timings by considering the simultaneity window. These advancements will further enhance the potential of MoHeat to deliver realistic and emotionally resonant thermal feedback in interactive applications.

ACKNOWLEDGEMENT

This work is supported in part by Suntory Holdings Limited, JST Moonshot R&D Program Grant Number JPMJMS2292, JSPS KAKENHI JP24K23883, JP24K02969 and JP24K22316.